\documentclass[
    aps,
    prl,
    twocolumn,
    nofootinbib,  
    floatfix,
    amsmath,
    superscriptaddress,
]{revtex4-2}

\usepackage{graphicx}
\usepackage{dcolumn}
\usepackage{bm}
\usepackage{hyperref}
\usepackage{todonotes}
\usepackage{placeins}
\usepackage{comment}
\usepackage{float}
\usepackage{booktabs}
\usepackage{overpic}
\usepackage{calc} 
\usepackage[normalem]{ulem}

\newcommand{\np}{1$e$N$p$0$\pi$ }
\newcommand{\zp}{1$e$0$p$0$\pi$ }

\begin{document}

\title{Search for an Anomalous Production of Charged-Current $\nu_e$ Interactions Without Visible Pions Across Multiple Kinematic Observables in MicroBooNE}

\newcommand{\ANL}{Argonne National Laboratory (ANL), Lemont, IL, 60439, USA}
\newcommand{\Bern}{Universit{\"a}t Bern, Bern CH-3012, Switzerland}
\newcommand{\BNL}{Brookhaven National Laboratory (BNL), Upton, NY, 11973, USA}
\newcommand{\UCSB}{University of California, Santa Barbara, CA, 93106, USA}
\newcommand{\Cambridge}{University of Cambridge, Cambridge CB3 0HE, United Kingdom}
\newcommand{\CIEMAT}{Centro de Investigaciones Energ\'{e}ticas, Medioambientales y Tecnol\'{o}gicas (CIEMAT), Madrid E-28040, Spain}
\newcommand{\Chicago}{University of Chicago, Chicago, IL, 60637, USA}
\newcommand{\Cincinnati}{University of Cincinnati, Cincinnati, OH, 45221, USA}
\newcommand{\CSU}{Colorado State University, Fort Collins, CO, 80523, USA}
\newcommand{\Columbia}{Columbia University, New York, NY, 10027, USA}
\newcommand{\Edinburgh}{University of Edinburgh, Edinburgh EH9 3FD, United Kingdom}
\newcommand{\FNAL}{Fermi National Accelerator Laboratory (FNAL), Batavia, IL 60510, USA}
\newcommand{\Granada}{Universidad de Granada, Granada E-18071, Spain}
\newcommand{\IIT}{Illinois Institute of Technology (IIT), Chicago, IL 60616, USA}
\newcommand{\ICL}{Imperial College London, London SW7 2AZ, United Kingdom}
\newcommand{\Indiana}{Indiana University, Bloomington, IN 47405, USA}
\newcommand{\KSU}{Kansas State University (KSU), Manhattan, KS, 66506, USA}
\newcommand{\Lancaster}{Lancaster University, Lancaster LA1 4YW, United Kingdom}
\newcommand{\LANL}{Los Alamos National Laboratory (LANL), Los Alamos, NM, 87545, USA}
\newcommand{\Louisiana}{Louisiana State University, Baton Rouge, LA, 70803, USA}
\newcommand{\Manchester}{The University of Manchester, Manchester M13 9PL, United Kingdom}
\newcommand{\MIT}{Massachusetts Institute of Technology (MIT), Cambridge, MA, 02139, USA}
\newcommand{\Michigan}{University of Michigan, Ann Arbor, MI, 48109, USA}
\newcommand{\MSU}{Michigan State University, East Lansing, MI 48824, USA}
\newcommand{\Minnesota}{University of Minnesota, Minneapolis, MN, 55455, USA}
\newcommand{\Nankai}{Nankai University, Nankai District, Tianjin 300071, China}
\newcommand{\NMSU}{New Mexico State University (NMSU), Las Cruces, NM, 88003, USA}
\newcommand{\Oxford}{University of Oxford, Oxford OX1 3RH, United Kingdom}
\newcommand{\Pitt}{University of Pittsburgh, Pittsburgh, PA, 15260, USA}
\newcommand{\QMUL}{Queen Mary University of London, London E1 4NS, United Kingdom}
\newcommand{\Rutgers}{Rutgers University, Piscataway, NJ, 08854, USA}
\newcommand{\SLAC}{SLAC National Accelerator Laboratory, Menlo Park, CA, 94025, USA}
\newcommand{\SDSMT}{South Dakota School of Mines and Technology (SDSMT), Rapid City, SD, 57701, USA}
\newcommand{\Maine}{University of Southern Maine, Portland, ME, 04104, USA}
\newcommand{\Syracuse}{Syracuse University, Syracuse, NY, 13244, USA}
\newcommand{\TelAviv}{Tel Aviv University, Tel Aviv, Israel, 69978}
\newcommand{\UTA}{University of Texas, Arlington, TX, 76019, USA}
\newcommand{\Tufts}{Tufts University, Medford, MA, 02155, USA}
\newcommand{\VTech}{Center for Neutrino Physics, Virginia Tech, Blacksburg, VA, 24061, USA}
\newcommand{\Warwick}{University of Warwick, Coventry CV4 7AL, United Kingdom}

\affiliation{\ANL}
\affiliation{\Bern}
\affiliation{\BNL}
\affiliation{\UCSB}
\affiliation{\Cambridge}
\affiliation{\CIEMAT}
\affiliation{\Chicago}
\affiliation{\Cincinnati}
\affiliation{\CSU}
\affiliation{\Columbia}
\affiliation{\Edinburgh}
\affiliation{\FNAL}
\affiliation{\Granada}
\affiliation{\IIT}
\affiliation{\ICL}
\affiliation{\Indiana}
\affiliation{\KSU}
\affiliation{\Lancaster}
\affiliation{\LANL}
\affiliation{\Louisiana}
\affiliation{\Manchester}
\affiliation{\MIT}
\affiliation{\Michigan}
\affiliation{\MSU}
\affiliation{\Minnesota}
\affiliation{\Nankai}
\affiliation{\NMSU}
\affiliation{\Oxford}
\affiliation{\Pitt}
\affiliation{\QMUL}
\affiliation{\Rutgers}
\affiliation{\SLAC}
\affiliation{\SDSMT}
\affiliation{\Maine}
\affiliation{\Syracuse}
\affiliation{\TelAviv}
\affiliation{\UTA}
\affiliation{\Tufts}
\affiliation{\VTech}
\affiliation{\Warwick}

\author{P.~Abratenko} \affiliation{\Tufts}
\author{D.~Andrade~Aldana} \affiliation{\IIT}
\author{L.~Arellano} \affiliation{\Manchester}
\author{J.~Asaadi} \affiliation{\UTA}
\author{A.~Ashkenazi}\affiliation{\TelAviv}
\author{S.~Balasubramanian}\affiliation{\FNAL}
\author{B.~Baller} \affiliation{\FNAL}
\author{A.~Barnard} \affiliation{\Oxford}
\author{G.~Barr} \affiliation{\Oxford}
\author{D.~Barrow} \affiliation{\Oxford}
% ...

\collaboration{The MicroBooNE Collaboration}
\thanks{microboone\_info@fnal.gov}\noaffiliation

\date{\today}

\begin{abstract}
This Letter presents an investigation of low-energy electron-neutrino interactions in the Fermilab Booster Neutrino Beam by the MicroBooNE experiment, motivated by the excess of electron-neutrino-like events observed by the MiniBooNE experiment.  This is the first measurement to use data from all five years of operation of the MicroBooNE experiment, corresponding to an exposure of $1.11\times 10^{21}$ protons on target, a 70\% increase on past results.  Two samples of electron neutrino interactions without visible pions are used, one with visible protons and one without any visible protons. The MicroBooNE data show reasonable agreement with the nominal prediction, with $p$-values $\ge 26.7\%$ when the two $\nu_e$ samples are combined, though the prediction exceeds the data in limited regions of phase space. The data is further compared to two empirical models that modify the predicted rate of  electron-neutrino interactions in different variables in the simulation to match the unfolded MiniBooNE low energy excess. In the first model, this unfolding is performed as a function of electron neutrino energy, while the second model aims to match the observed shower energy and angle distributions of the MiniBooNE excess. This measurement excludes an electron-like interpretation of the MiniBooNE excess based on these models at $> 99\%$ CL$_\mathrm{s}$ in all kinematic variables.

\end{abstract}

\maketitle
\emph{Introduction.---}The low energy excess (LEE) of electromagnetic activity observed by the MiniBooNE experiment~\cite{MiniBooNE:2018esg} is one of the most puzzling anomalous results in neutrino physics to date. Together with the LSND anomaly \cite{PhysRevD.64.112007}, it may hint at neutrino oscillations that cannot be accommodated by the three known neutrino flavors, or other new physics.
The MicroBooNE experiment~\cite{MicroBooNE:2016pwy} has been designed to determine the nature of this anomaly.  For this purpose, it operates in the same Booster Neutrino Beamline (BNB) at Fermilab as the MiniBooNE experiment and in the same $\mathcal{O}$(1 GeV) energy range. It is located at a baseline of $470$~m from the beam target. In contrast to MiniBooNE, the liquid argon time projection chamber technology used in MicroBooNE is capable of tracking all energetic charged particles produced in neutrino interactions with a spatial resolution of around 2--3~mm, and therefore allows a detailed separation of observed events by their final state topology. A set of measurements using the first three years of MicroBooNE data explored the possibility of an excess due to an electron neutrino rate enhancement in multiple topologies~\cite{MicroBooNE:2021wad,MicroBooNE:2021tya,MicroBooNE:2021pvo,MicroBooNE:2021nxr}, or an enhancement in the rate of neutral current produced Delta-baryon decays producing photons~\cite{MicroBooNE:2021zai}. Other theoretical models have been proposed to explain the MiniBooNE LEE. These include Standard Model backgrounds~\cite{MiniBooNE:2020pnu,MicroBooNE:2021zai,Kelly:2022uaa,Brdar_2022}, light sterile neutrinos (and variations)~\cite{Diaz:lightsterile,Sebas:lightsterile,Vergani:2021tgc}, heavy neutrino decay~\cite{Alvarez-Ruso:2017hdm,Fischer:heavysterile}, and dark-sector particles~\cite{Abdullahi:2020nyr,Bertuzzo:2018itn,Ballett:2018ynz,Bertuzzo:2018itn}.



The analysis presented in this Letter builds on our previous result~\cite{MicroBooNE:2021wad} and adds data collected from MicroBooNE’s full five-year operation span. As in the earlier analysis, this analysis measures charged current (CC) $\nu_e$ interactions without pions in the final state, mimicking the signature of the MiniBooNE excess. We further sub-divide this selection into separate channels for events with and without protons in the final state, which the MiniBooNE detector could not differentiate. This updated analysis employs the same signal definition, Monte Carlo (MC) simulation (flux, GENIE neutrino interaction model~\cite{GenieUBTune}, particle transport model~\cite{ALLISON2016186, Geant:1610988, AGOSTINELLI2003250}, and detector simulation), as well as the same event reconstruction~\cite{pandora}. This update brings several major improvements to the analysis. First is the inclusion of data collected between 2018 and 2020, leading to a 70\% larger dataset corresponding to $1.11\times 10^{21}$ protons on target (POT). This follows extensive validations of detector stability in the new dataset. A new model for the MiniBooNE excess using visible shower energy and angle is also introduced to better investigate this anomaly across multiple kinematic variables, together with an expanded set of constraint channels used to better constrain the intrinsic $\nu_e$ rate in the detector as well as $\pi^0$ backgrounds, and an improved treatment of detector systematic uncertainties is adopted. Finally, this analysis makes expanded use of the Cosmic Ray Tagger (CRT)~\cite{bib:CRT} system installed in the detector in 2017. The expanded dataset used for this work has larger overlap with the period in which the CRT was operational, improving the analysis’s ability to reject cosmic ray backgrounds.

\emph{Event selection.---}We consider two signal channels in this analysis: The \zp channel targets CC $\nu_e$ interactions with no visible protons or pions in the final state, while the \np channel modifies this by demanding at least one proton with kinetic energy above 40 MeV. This threshold is selected to correspond to an average propagation distance of approximately 1~cm. The presence of a proton at the neutrino interaction vertex for \np events enhances the selection efficiency and reduces the background of this channel compared to the \zp channel. 
The \np and \zp selection procedures were not changed from those presented in Ref.~\cite{MicroBooNE:2021wad}, except for the addition of new CRT information to boost the background rejection in the \zp channel. Over the full five-year dataset, the use of the CRT system rejects an additional 25.4\% of the remaining cosmic ray background in the \zp channel, while retaining 98.9\% of the signal surviving the cuts already applied. Standard Model contributions to both signal channels are dominated by CC electron neutrinos, with neutral current (NC) interactions containing final-state $\pi^0$ mesons constituting the dominant background, particularly in the \zp channel.


\begin{figure}
    \centering
    \includegraphics[width=\linewidth]{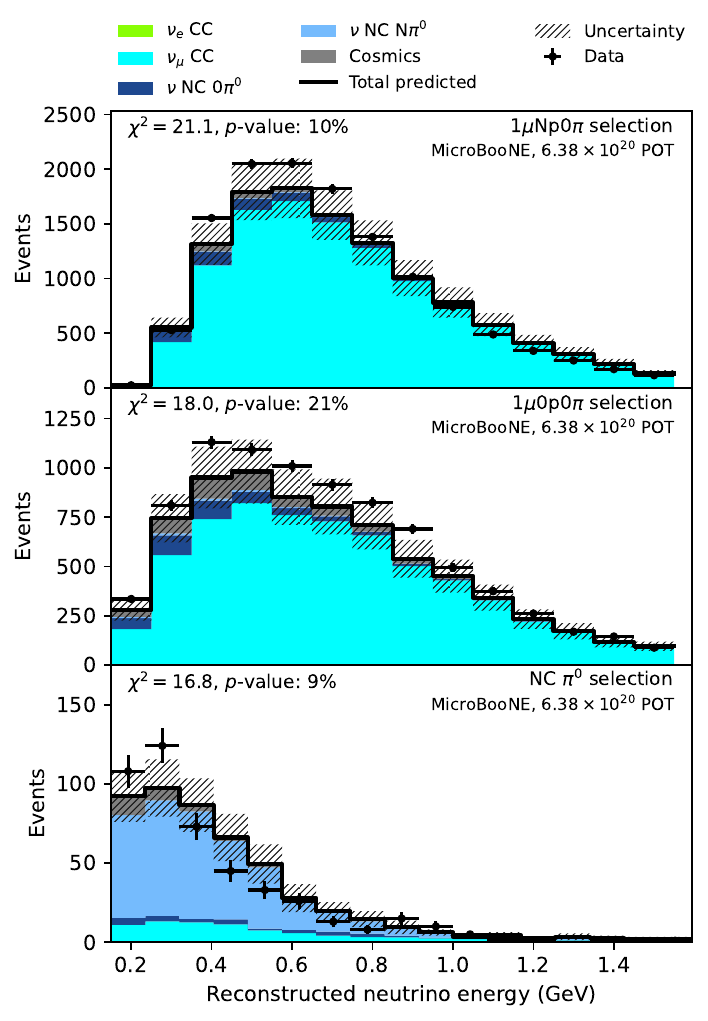}
    \caption{Distributions of control sample events used in this analysis. The prediction is broken down into CC $\nu_e$ and $\nu_{\mu}$ interactions, NC interactions not producing neutral pions, NC interactions that produce neutral pions, and cosmic rays mistaken for neutrino interactions. Only bins up to 1~GeV from the  NC $\pi^0$ selection are used in the constraint procedure due to low statistics above this energy.}
    \label{fig:sideband-selections}
\end{figure}

This analysis uses control samples in the data to constrain the predicted event rates and systematic uncertainties in the \np and \zp  channels in a data-driven way. Two of these control channels employ CC $\nu_{\mu}$ interactions in the absence of pions, 1$\mu$N$p$0$\pi$ and 1$\mu$0$p$0$\pi$, matching the final state hadronic observables of the electron neutrino samples. The high-statistics $\nu_{\mu}$ channels aim to constrain the $\nu_e$ predictions, leveraging strong correlations in cross-section (through lepton universality), flux (through their shared neutrino parentage from pions and kaons decaying in the BNB), and detector uncertainties. 
A third sample is designed to constrain the prediction for events with mis-reconstructed neutral pions in the \zp signal channel.
The selection identifies events with at least two reconstructed electromagnetic showers and no visible muons, protons, or charged pions to enhance the number of NC interactions containing $\pi^{0}$ mesons (referred to as $\nu$ NC $\pi^{0}$ events).  This requirement is orthogonal to the electron neutrino selection, which requires exactly one reconstructed electromagnetic shower to limit the $\pi^{0}$ meson background.

The distributions of observed events in the three control samples are shown in Fig.~\ref{fig:sideband-selections}. The control samples are binned in reconstructed neutrino energy when calculating their covariance with the predictions in the signal channels, regardless of the variable used in the signal distribution. The reconstructed neutrino energy is defined as the sum of the energies of all reconstructed electromagnetic showers and charged particle tracks. Shower energies are reconstructed based on calorimetric information \cite{Adams_2020}, while track energies are based on the range \cite{nist_stopping_power,pdg_muon_stopping_power} using ionization-energy-loss measurements to assign a proton or muon identity to each track \cite{MicroBooNE:2021ddy}. The frequentist $p$-values of the data given the statistical and systematic uncertainties are 10\% and 21\% for the 1$\mu$N$p$0$\pi$ and 1$\mu$0$p$0$\pi$ channels, respectively, while the $p$-value for the NC~$\pi^0$ channel is 9.1\%, indicating agreement within uncertainties. The control samples use data collected during periods when information from the CRT was available, corresponding to $6.4 \times 10^{20}$ POT, a factor of three more than used in our previous result \cite{MicroBooNE:2021wad}.


\emph{Systematic uncertainties.---}The set of systematic uncertainties applied in this analysis is unchanged from our previous result~\cite{MicroBooNE:2021wad}. Four categories are included: neutrino flux, neutrino interaction cross sections, secondary interactions of hadrons outside of the target nucleus, and the detector response model. The covariance matrices for each of these sources of uncertainties are calculated including the correlations between all signal and control samples and added together. Of the included categories of uncertainty, neutrino cross sections are the largest source of systematic uncertainties in this analysis providing on average 50\% of the total variance on the expectation value in each bin.

In the case of the flux, cross section, and secondary interaction uncertainties, variations from the central value are obtained through a reweighting procedure, while for the detector response uncertainties the covariance is obtained by propagating a set of simulated neutrino interactions through several detector models \cite{MicroBooNE:2021roa}. As an update to the earlier iteration of this analysis, correlations in detector systematics are now taken into account. In order to limit the effect of statistical fluctuations in the detector variation samples on the bin-to-bin correlations, we smooth the histograms for each sample by convolving the bin counts with a pseudo-gaussian filter before calculating the covariance matrix. The inclusion of the correlation terms in the detector systematics covariance matrix allows these uncertainties to be constrained by the $\nu_{\mu}$ data, which reduces the overall systematic uncertainty on the signal after constraints by 17\%. Further details on this smoothing procedure and its effect on the covariance matrix, as well as a detailed breakdown of the contribution of each source of uncertainty to the overall systematic errors, are given in the supplemental material~\cite{SupplementalMaterial}.

\emph{Sideband constraint.---}From the set of variations of all the systematic uncertainties under consideration, we calculate the matrix describing the covariances between the predicted event rates in the histograms of the two signal channels and three control channels described above. We then use the block matrix method \cite{Eaton1983-xi} to constrain the prediction in the \np and \zp signal channels using the data in the $\nu_\mu$ and $\pi^0$ control channels. This mathematical procedure modifies the prediction in the signal channels based on the bin-to-bin difference between the data and predictions in the control channels and the uncertainties and correlations across the signal and control channels. The constraint is applied to LEE and background predictions, where the LEE is treated in the same way as the electron-neutrinos selected. On average, the constraint procedure reduces the uncertainty on the predicted event rates by 40\%. The predictions in the \np and \zp channels are mostly increased after the application of this constraint due to positive correlations between the $\nu_e$ and $\nu_{\mu}$ predictions and the underpredictions shown in Fig.~\ref{fig:sideband-selections}. The full covariance matrix and the impact of these constraints on the predicted event rate and systematic uncertainty are illustrated in the supplemental material~\cite{SupplementalMaterial}.

The statistical uncertainty due to Poisson fluctuations of the data is incorporated into the analysis by adding a combined Neyman-Pearson term~\cite{Ji:2019yca} to the covariance matrix obtained from systematic variations. The resulting total covariance matrix is used to calculate a $\chi^2$ statistic between the predicted and observed bin counts. On average, the Poisson uncertainties are three times as large as the systematic uncertainties after constraints.

\emph{Signal model.---} All models attempting to explain the LEE ultimately rely on assumptions about the interaction process that produces the observed excess events in the MiniBooNE detector.
For this analysis, we build two empirical models which aim to test the electron-like LEE hypothesis in three kinematic variables: neutrino energy, electron energy, and electron angle $\theta$. These are the three variables reported in MiniBooNE results \cite{MiniBooNE:2018esg,PhysRevLett.98.231801,MiniBooNE:2020pnu}. The choice to add the two electron kinematic variables in this analysis is motivated by the fact that these are the direct observables that the MiniBooNE detector measures, and any model suggested to explain the LEE must inevitably be consistent with the observations in these variables. Furthermore, the phenomenological work produced in recent years to explain the MiniBooNE excess mentioned above relies heavily on these kinematics, making them an interesting and natural choice. The inclusion of such variables in this work elevates the impact and interpretability of the results investigating the MiniBooNE excess. These two models are to be considered as benchmark models, and for simplicity they neglect systematic uncertainty on the MiniBooNE LEE signal prediction.

We define \textit{LEE Signal Model 1} as the same $\nu_e$-like LEE model tested in Ref.~\cite{MicroBooNE:2021wad}.
This model assumes that the LEE is entirely originated from an energy-dependent enhancement of MiniBooNE's intrinsic $\nu_e$ flux and is generated by unfolding the observed MiniBooNE excess as a function of reconstructed neutrino energy using the smearing matrix that describes the relationship between the true and reconstructed energy of CC quasi-elastic events in the MiniBooNE detector.
This model suffers the limitation of producing shower kinematics that do not align with the shower energy and angle measurements observed by MiniBooNE, particularly for forward-going showers~\cite{MiniBooNE:2020pnu}.

To overcome this limitation, we construct a new signal model -- labelled \textit{LEE Signal Model 2} --  by unfolding the background-subtracted excess of data events in the two-dimensional space of reconstructed shower energy and $\cos(\theta)$ as shown in the supplemental material~\cite{SupplementalMaterial}. This model does not assume an underlying origin of the excess events but aims to reflect the excess in MiniBooNE's observed final-state electrons. The unfolding process utilizes MiniBooNE's selection efficiency of final-state electrons and the matrix that describes the smearing of reconstructed electron kinetic energy and angle. The ratio between the background-subtracted unfolded excess data rate in electron kinematics and the corresponding predicted rate from MiniBooNE's intrinsic $\nu_e$ flux from the BNB without an excess is used to obtain a scaling in each bin of electron energy and angle. This scaling is then applied to the true electron kinematics from MicroBooNE's intrinsic $\nu_e$ prediction to generate the predicted signal, leaving the modeling of the hadronic kinematics unchanged. More information on \textit{LEE Signal Model 2} can be found in the supplemental material~\cite{SupplementalMaterial}.

\begin{figure}
    \centering
    \includegraphics[width=\linewidth]{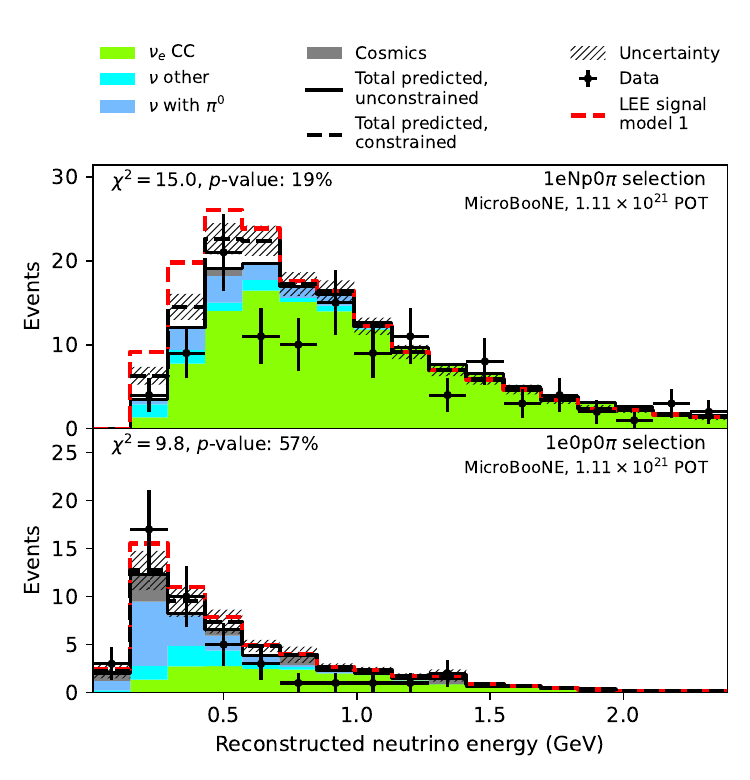}
     \caption{Distribution of MC simulation compared with data for reconstructed neutrino energy in the \np and \zp signal channels, along with the \textit{LEE Signal Model 1}. Statistical tests only take bins between 0.15~GeV and 1.55~GeV into account.}
     \label{fig:old-signal-nu-e}
\end{figure}

\begin{figure}
    \centering
    \includegraphics[width=\linewidth]{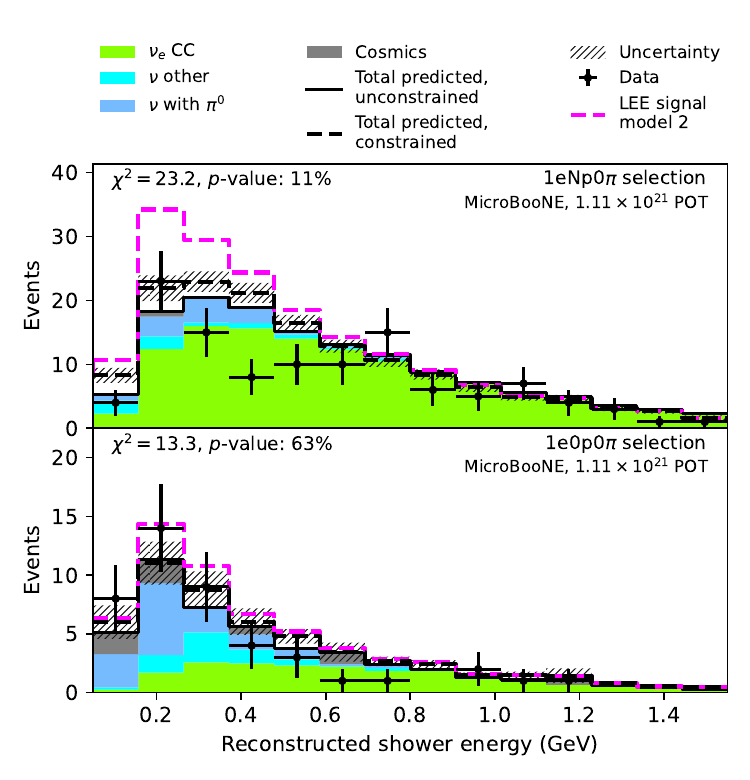}
     \caption{Distribution of MC simulation compared with data for reconstructed shower energy in the \np and \zp signal channels, along with the \textit{LEE Signal Model 2}.}
     \label{fig:new-signal-shr-e}
\end{figure}

\begin{figure}
    \centering
    \includegraphics[width=\linewidth]{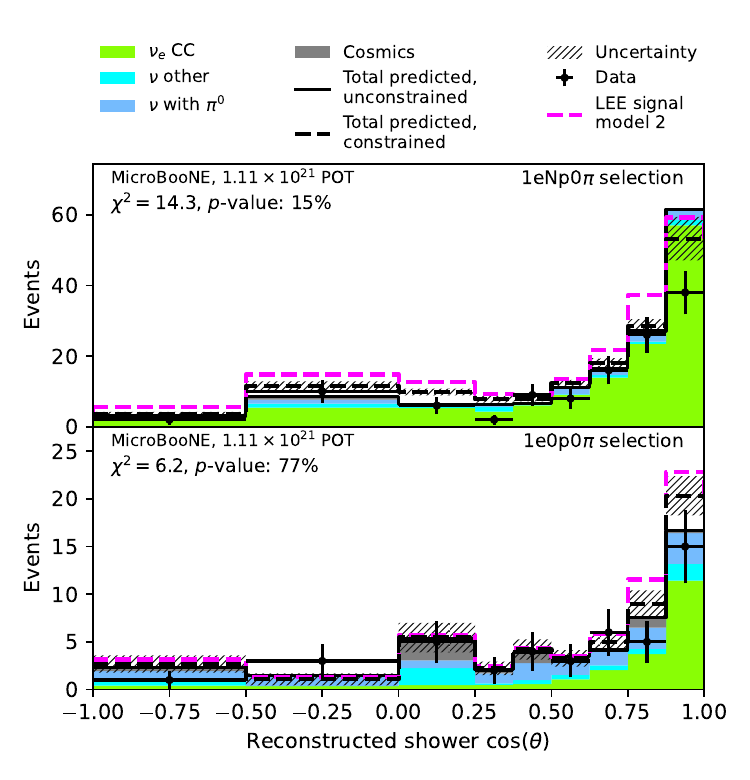}
     \caption{Distribution of MC simulation compared with data for reconstructed shower $\cos(\theta)$ in the \np and \zp signal channels, along with the \textit{LEE Signal Model 2}.}
     \label{fig:new-signal-shr-costheta}
\end{figure}

\emph{Results.---} In addition to the stability checks performed on the new data, an extensive set of validations using selections excluding the signal regions were performed to test the modeling of selection variables prior to unblinding of the signal distributions. The final distributions of data and MC simulation are presented as a function of reconstructed neutrino energy for both channels in Fig.~\ref{fig:old-signal-nu-e} with the prediction for the \textit{LEE Signal Model 1}. In this and subsequent $\nu_e$ spectra shown, solid stacked histograms show the unconstrained predicted event rate for various event topologies, while the total constrained prediction is shown with a dashed black line. Some over-prediction can be observed at medium energies in the \np channel, which has also been seen in the previous analysis~\cite{MicroBooNE:2021wad}.
With respect to our previous result, the ratio between data and MC in these bins has moved closer to unity while the uncertainties have decreased, leaving the $\chi^2$ in the \np channel nearly identical at a $p$-value of 19\%. The agreement between data and simulation in the \zp channel has improved from a $p$-value of 13\% in the previous analysis to 57\% in this work; in particular the tension between data and MC that was seen in the second bin is now gone.
Distributions for shower kinematics and the prediction for the \textit{LEE Signal Model 2} are displayed in Figs.~\ref{fig:new-signal-shr-e} - \ref{fig:new-signal-shr-costheta}. These show some over-prediction at intermediate electron energies and forward angles in the \np channel; the prediction and the observation match well in the \zp channel. A similar behavior has previously been seen in an electron-neutrino cross-section measurement performed by MicroBooNE~\cite{MicroBooNE:2022tdd}.

\begin{figure}
\vspace{0.3cm}
    \centering
    \includegraphics[width=\linewidth]{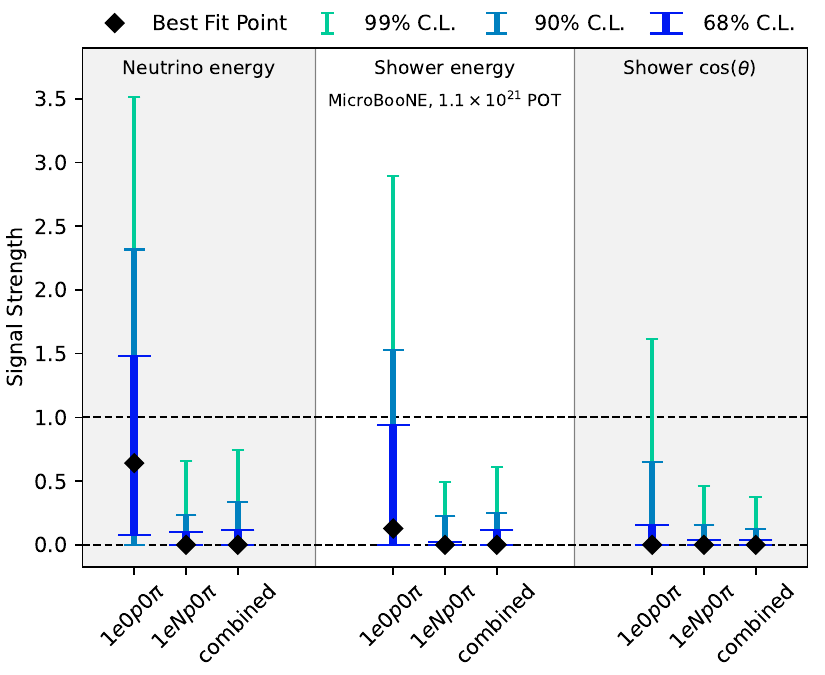}
    \caption{Confidence intervals obtained from all fits to the LEE signal strength performed in this analysis, with 0 corresponding to no excess and 1 corresponding to the size of the signal observed by MiniBooNE. Confidence intervals shown are generated with the Feldman-Cousins procedure~\cite{Feldman:1997qc}.}
    \label{fig:ConfidenceIntervals}
\end{figure}

For both models, three statistical tests are performed on the data: a simple $\chi^2$ test assuming no excess over the standard model ($H_0$, the null hypothesis), a $\Delta \chi^2$ test comparing $H_0$ to the nominal MiniBooNE excess ($H_1$, the alternative hypothesis) via the test statistic $\Delta \chi^2 = \chi^2_{H_0} - \chi^2_{H_1}$, and finally a fit for the strength of the MiniBooNE signal, assumed as a overall scaling parameter. In every statistical test, we calculate frequentist $p$-values using pseudo-data trials. For each trial, we first sample a histogram of expectation values from a multivariate normal distribution according to the covariance matrix of systematic uncertainties and then draw the pseudo-data from Poisson distributions with these expectation values. The tests are performed separately on the neutrino energy, shower energy, and shower $\cos(\theta)$ distributions. In the case of the neutrino energy distributions shown in Fig.~\ref{fig:old-signal-nu-e}, the histogram for the LEE hypothesis is calculated using \emph{LEE Signal Model 1}. This calculation is described in more detail in Section IIID of Ref.~\cite{MicroBooNE:2021wad}. The use of the same binning and signal model makes the results from this channel directly comparable to Ref. \cite{MicroBooNE:2021wad}. Although \textit{LEE Signal Model 2} is defined in two dimensions, we perform the statistical tests independently on the one-dimensional projections of the data onto the shower energy and shower angle variables, since the data statistics are insufficient to fill a two-dimensional histogram. The $p$-values of these tests are therefore not entirely independent. The outcomes of all statistical tests are summarized in Table~\ref{tab:SummaryRes}.
\begin{table*}
\caption{Results with data corresponding to $1.11\times10^{21}$ POT. The first three rows show the $\chi^2$ between the data and the null hypothesis after constraint ($H_0$) and its corresponding $p$-value. Rows 4 through 8 show the results of the two-hypothesis test in which $H_0$ is compared to the signal model hypotheses ($H_1$). 
The median sensitivity gives the confidence level at which we would be able to reject the null hypothesis at the median $\Delta \chi^2$ expected under $H_1$.
Finally, the confidence level for rejecting $H_1$ using the CL$_{\mathrm{s}}$ method is reported. The last three rows show the best fit point of the fitted signal strength, $\mu_\mathrm{BF}$, its upper limit at $2\sigma$ C.L. and the expected upper limit for the case that the data corresponded exactly to the prediction at $H_0$. \label{tab:SummaryRes}}
\centering

\begin{tabular}{ r | c  c  c | c  c  c | c  c  c | c }
Signal Model $(H_1)$ & \multicolumn{3}{c|}{LEE Signal Model 1} &\multicolumn{6}{c|}{LEE Signal Model 2} \\
Variable & \multicolumn{3}{c|}{Neutrino Energy} & \multicolumn{3}{c|}{Electron Energy} & \multicolumn{3}{c|}{Electron $\cos(\theta$)}  \\
Channel & \np & \zp & Combined & \np & \zp & Combined & \np & \zp & Combined & Row \\
 \hline
Observed $\chi^2$ & 14.9& 9.8& 24.8& 23.2& 13.3& 35.8& 14.3& 6.2 & 19.8 & 1\\
ndof & 10 & 10 & 20 & 14 & 14 & 28 & 9 & 9 & 18 & 2 \\
$P(\chi^2 > $obs.$ | H_0)$ [\%] & 18.6& 57.1& 32.0& 10.8& 62.7& 26.7& 15.3& 77.2& 43.6& 3\\
 \hline
obs. $H_0 - H_1$ $\Delta\chi^2$ & -11.4& 0.4 & -10.4& -15.3& -1.2 & -15.2& -17.5& -4.6 & -20.6& 4\\
$P(\Delta\chi^2$$<$obs.$|H_0$) [\%] & 5.9 & 78.7& 16.6& 10.0 & 59.5& 17.3 & 2.0  & 12.4& 1.32& 5\\
$P(\Delta\chi^2$$<$obs.$|H_1$) [\%] & 0.04& 33.6& 0.07& 0.002& 13.7& 0.003& 0.001& 0.8 & 0.0002& 6\\
Median sensitivity [\%] & 1.12& 11.0 & 0.54& 0.06 & 8.6& 0.02& 0.46 & 8.4 & 0.20& 7\\

1 - $\mathrm{CL}_\mathrm{s}$ [\%] & 0.59& 43& 0.41&  0.017&23&0.02&0.05&6.7& 0.014& 8\\
 \hline
$\mu_\texttt{BF}$ & 0.00 & 0.61 & 0.00 & 0.00 & 0.14& 0.00 & 0.00 & 0.00 & 0.00& 9 \\
2$\sigma$ C.L. upper limit on $\mu$ & 0.35& 2.63& 0.50& 0.33 & 1.89& 0.38& 0.26& 0.87 & 0.22& 10\\
Exp. 2$\sigma$ C.L. limit & 1.02& 1.89& 0.88& 0.71& 1.81& 0.64& 0.84& 1.82 & 0.75& 11 
\end{tabular}
\end{table*}

The frequentist $p$-values of the observed data under the null-hypothesis are 32\%, 27\%, and 44\% for the combined signal channels in the neutrino energy, shower energy, and shower $\cos(\theta)$ distributions, respectively. When looking at the separate channels, $p$-values are all $>$ 50\% in the \zp channel and $>$10\% in the \np channel,  regardless of the variable. For the \np channel, this indicates that these distributions are compatible with the prediction after constraint at the 
1.6$\sigma$ level with no LEE signal added. When all \np events in the energy range between 0.15~GeV and 1.55~GeV are collapsed into a single bin, the constrained prediction at the null hypothesis is 133.5 $\pm$ 7.4 (syst.) events while the observed bin count is 102 events. This corresponds to a deficit of 24\% with a statistical significance of 2.4$\sigma$ in this channel. Past~\cite{MicroBooNE:2022tdd,PhysRevD.105.L051102} as well as ongoing measurements of $\nu_e$ cross-sections on argon by MicroBooNE may help elucidate this finding, and interpretations of MicroBooNE’s $\nu_e$ data under a sterile neutrino oscillation $\nu_e$ disappearance hypothesis can be found in MicroBooNE's dedicated search~\cite{MicroBooNE:2022sdp}.

In order to account for the over-prediction in the \np channel, in the two-hypothesis test we employ the modified frequentist CL$_{\mathrm{s}}$ method that has been proposed in Ref.~\cite{CLs_method, Read:451614} and that yields more conservative rejection levels with under-fluctuating data. Using this approach, we reject
the \textit{LEE Signal Model 1} at $2.9\sigma$ and the \textit{LEE Signal Model 2} at $3.7\sigma$ and $3.8\sigma$ when measured in the electron energy or electron angle, respectively. Further details on the two-hypothesis tests performed are provided in the supplemental material~\cite{SupplementalMaterial}.

The results of the signal strength fits for all three measured variables, and for each individual signal channel as well as their combination, are shown in Fig.~\ref{fig:ConfidenceIntervals}, where we have used the Feldman-Cousins \cite{Feldman:1997qc} procedure to calculate frequentist confidence intervals. When fitting to both channels, the LEE hypothesis is excluded at $>99\%$ confidence level (C.L.) in all three variables. The two-hypothesis tests and the signal strength fits for measurements using only the \zp signal channel show a weaker preference for $H_0$  over $H_1$. This demonstrates that the strong exclusion of the LEE hypothesis in the combined fit is mostly driven by the \np channel. These results are limited by statistical uncertainties; while this measurement includes data from all MicroBooNE run periods, reconstruction and analysis updates could improve the efficiency and reduce the statistical uncertainty in future electron neutrino analyses.

\emph{Summary.---}This Letter presents an updated investigation of low energy electron-like events from the BNB using an expanded MicroBooNE dataset of $1.11\times 10^{21}$ POT. The analysis builds on earlier measurements which test the hypothesis that an anomalous excess is due to an energy-dependent increase in the intrinsic $\nu_e$ rate and complements this with new tests motivated by the observable kinematics (shower angle and energy) reported by MiniBooNE. For \np and \zp events, in all kinematic variables, we find that our data is not consistent with an excess of $\nu_e$ interactions, with $p$-values showing agreement with the null hypothesis in the 10$\%$--77$\%$ range, depending on the final state and the kinematic variable considered. Both signal models tested in this measurement that interpret the MiniBooNE LEE as $\nu_e$ events are excluded at $\geq$ 99\% CL$_{\mathrm{s}}$. While the MiniBooNE excess remains unexplained, our measurements confirm and strengthen our previous results. By exploring a more comprehensive set of variables to test the LEE and leveraging an increased statistics dataset, we find that MicroBooNE data is inconsistent with an electron-like interpretation of the MiniBooNE LEE.

\emph{Acknowledgments.---}This document was prepared by the MicroBooNE Collaboration using the resources of the Fermi National Accelerator Laboratory (Fermilab), a U.S. Department of Energy, Office of Science, HEP User Facility. Fermilab is managed by Fermi Research Alliance, LLC (FRA), acting under Contract No. DE-AC02-07CH11359. MicroBooNE is supported by the following: the U.S. Department of Energy, Office of Science, Offices of High Energy Physics and Nuclear Physics; the U.S. National Science Foundation; the Swiss National Science Foundation; the Science and Technology Facilities Council (STFC), part of the United Kingdom Research and Innovation; the Royal Society (United Kingdom); and The European Union’s Horizon 2020 Marie Skłodowska-Curie Actions. Additional support for the laser calibration system and cosmic ray tagger was provided by the Albert Einstein Center for Fundamental Physics, Bern, Switzerland. We also acknowledge the contributions of technical and scientific staff to the design, construction, and operation of the MicroBooNE detector as well as the contributions of past collaborators to the development of MicroBooNE analyses, without whom this work would not have been possible.

\bibliographystyle{apsrev4-2}
\bibliography{refs}

\begin{thebibliography}{41}%
\makeatletter
\providecommand \@ifxundefined [1]{%
 \@ifx{#1\undefined}
}%
\providecommand \@ifnum [1]{%
 \ifnum #1\expandafter \@firstoftwo
 \else \expandafter \@secondoftwo
 \fi
}%
\providecommand \@ifx [1]{%
 \ifx #1\expandafter \@firstoftwo
 \else \expandafter \@secondoftwo
 \fi
}%
\providecommand \natexlab [1]{#1}%
\providecommand \enquote  [1]{``#1''}%
\providecommand \bibnamefont  [1]{#1}%
\providecommand \bibfnamefont [1]{#1}%
\providecommand \citenamefont [1]{#1}%
\providecommand \href@noop [0]{\@secondoftwo}%
\providecommand \href [0]{\begingroup \@sanitize@url \@href}%
\providecommand \@href[1]{\@@startlink{#1}\@@href}%
\providecommand \@@href[1]{\endgroup#1\@@endlink}%
\providecommand \@sanitize@url [0]{\catcode `\\12\catcode `\$12\catcode `\&12\catcode `\#12\catcode `\^12\catcode `\_12\catcode `\%12\relax}%
\providecommand \@@startlink[1]{}%
\providecommand \@@endlink[0]{}%
\providecommand \url  [0]{\begingroup\@sanitize@url \@url }%
\providecommand \@url [1]{\endgroup\@href {#1}{\urlprefix }}%
\providecommand \urlprefix  [0]{URL }%
\providecommand \Eprint [0]{\href }%
\providecommand \doibase [0]{https://doi.org/}%
\providecommand \selectlanguage [0]{\@gobble}%
\providecommand \bibinfo  [0]{\@secondoftwo}%
\providecommand \bibfield  [0]{\@secondoftwo}%
\providecommand \translation [1]{[#1]}%
\providecommand \BibitemOpen [0]{}%
\providecommand \bibitemStop [0]{}%
\providecommand \bibitemNoStop [0]{.\EOS\space}%
\providecommand \EOS [0]{\spacefactor3000\relax}%
\providecommand \BibitemShut  [1]{\csname bibitem#1\endcsname}%
\let\auto@bib@innerbib\@empty
\bibitem [{\citenamefont {Aguilar-Arevalo}\ \emph {et~al.}(2018)\citenamefont {Aguilar-Arevalo} \emph {et~al.}}]{MiniBooNE:2018esg}%
  \BibitemOpen
  \bibfield  {author} {\bibinfo {author} {\bibfnamefont {A.~A.}\ \bibnamefont {Aguilar-Arevalo}} \emph {et~al.} (\bibinfo {collaboration} {MiniBooNE Collaboration}),\ }\href {https://doi.org/10.1103/PhysRevLett.121.221801} {\bibfield  {journal} {\bibinfo  {journal} {Phys. Rev. Lett.}\ }\textbf {\bibinfo {volume} {121}},\ \bibinfo {pages} {221801} (\bibinfo {year} {2018})}\BibitemShut {NoStop}%
\bibitem [{\citenamefont {Aguilar}\ \emph {et~al.}(2001)\citenamefont {Aguilar} \emph {et~al.}}]{PhysRevD.64.112007}%
  \BibitemOpen
  \bibfield  {author} {\bibinfo {author} {\bibfnamefont {A.}~\bibnamefont {Aguilar}} \emph {et~al.} (\bibinfo {collaboration} {LSND Collaboration}),\ }\href {https://doi.org/10.1103/PhysRevD.64.112007} {\bibfield  {journal} {\bibinfo  {journal} {Phys. Rev. D}\ }\textbf {\bibinfo {volume} {64}},\ \bibinfo {pages} {112007} (\bibinfo {year} {2001})}\BibitemShut {NoStop}%
\bibitem [{\citenamefont {Acciarri}\ \emph {et~al.}(2017)\citenamefont {Acciarri} \emph {et~al.}}]{MicroBooNE:2016pwy}%
  \BibitemOpen
  \bibfield  {author} {\bibinfo {author} {\bibfnamefont {R.}~\bibnamefont {Acciarri}} \emph {et~al.} (\bibinfo {collaboration} {MicroBooNE Collaboration}),\ }\href {https://doi.org/10.1088/1748-0221/12/02/P02017} {\bibfield  {journal} {\bibinfo  {journal} {{J. Instrum.}}\ }\textbf {\bibinfo {volume} {12}},\ \bibinfo {pages} {P02017} (\bibinfo {year} {2017})}\BibitemShut {NoStop}%
\bibitem [{\citenamefont {Abratenko}\ \emph {et~al.}(2022{\natexlab{a}})\citenamefont {Abratenko} \emph {et~al.}}]{MicroBooNE:2021wad}%
  \BibitemOpen
  \bibfield  {author} {\bibinfo {author} {\bibfnamefont {P.}~\bibnamefont {Abratenko}} \emph {et~al.} (\bibinfo {collaboration} {MicroBooNE Collaboration}),\ }\href {https://doi.org/10.1103/PhysRevD.105.112004} {\bibfield  {journal} {\bibinfo  {journal} {Phys. Rev. D}\ }\textbf {\bibinfo {volume} {105}},\ \bibinfo {pages} {112004} (\bibinfo {year} {2022}{\natexlab{a}})}\BibitemShut {NoStop}%
\bibitem [{\citenamefont {Abratenko}\ \emph {et~al.}(2022{\natexlab{b}})\citenamefont {Abratenko} \emph {et~al.}}]{MicroBooNE:2021tya}%
  \BibitemOpen
  \bibfield  {author} {\bibinfo {author} {\bibfnamefont {P.}~\bibnamefont {Abratenko}} \emph {et~al.} (\bibinfo {collaboration} {MicroBooNE Collaboration}),\ }\href {https://doi.org/10.1103/PhysRevLett.128.241801} {\bibfield  {journal} {\bibinfo  {journal} {Phys. Rev. Lett.}\ }\textbf {\bibinfo {volume} {128}},\ \bibinfo {pages} {241801} (\bibinfo {year} {2022}{\natexlab{b}})}\BibitemShut {NoStop}%
\bibitem [{\citenamefont {Abratenko}\ \emph {et~al.}(2022{\natexlab{c}})\citenamefont {Abratenko} \emph {et~al.}}]{MicroBooNE:2021pvo}%
  \BibitemOpen
  \bibfield  {author} {\bibinfo {author} {\bibfnamefont {P.}~\bibnamefont {Abratenko}} \emph {et~al.} (\bibinfo {collaboration} {MicroBooNE Collaboration}),\ }\href {https://doi.org/10.1103/PhysRevD.105.112003} {\bibfield  {journal} {\bibinfo  {journal} {Phys. Rev. D}\ }\textbf {\bibinfo {volume} {105}},\ \bibinfo {pages} {112003} (\bibinfo {year} {2022}{\natexlab{c}})}\BibitemShut {NoStop}%
\bibitem [{\citenamefont {Abratenko}\ \emph {et~al.}(2022{\natexlab{d}})\citenamefont {Abratenko} \emph {et~al.}}]{MicroBooNE:2021nxr}%
  \BibitemOpen
  \bibfield  {author} {\bibinfo {author} {\bibfnamefont {P.}~\bibnamefont {Abratenko}} \emph {et~al.} (\bibinfo {collaboration} {MicroBooNE Collaboration}),\ }\href {https://doi.org/10.1103/PhysRevD.105.112005} {\bibfield  {journal} {\bibinfo  {journal} {Phys. Rev. D}\ }\textbf {\bibinfo {volume} {105}},\ \bibinfo {pages} {112005} (\bibinfo {year} {2022}{\natexlab{d}})}\BibitemShut {NoStop}%
\bibitem [{\citenamefont {Abratenko}\ \emph {et~al.}(2022{\natexlab{e}})\citenamefont {Abratenko} \emph {et~al.}}]{MicroBooNE:2021zai}%
  \BibitemOpen
  \bibfield  {author} {\bibinfo {author} {\bibfnamefont {P.}~\bibnamefont {Abratenko}} \emph {et~al.} (\bibinfo {collaboration} {MicroBooNE Collaboration}),\ }\href {https://doi.org/10.1103/PhysRevLett.128.111801} {\bibfield  {journal} {\bibinfo  {journal} {Phys. Rev. Lett.}\ }\textbf {\bibinfo {volume} {128}},\ \bibinfo {pages} {111801} (\bibinfo {year} {2022}{\natexlab{e}})}\BibitemShut {NoStop}%
\bibitem [{\citenamefont {Aguilar-Arevalo}\ \emph {et~al.}(2021)\citenamefont {Aguilar-Arevalo} \emph {et~al.}}]{MiniBooNE:2020pnu}%
  \BibitemOpen
  \bibfield  {author} {\bibinfo {author} {\bibfnamefont {A.~A.}\ \bibnamefont {Aguilar-Arevalo}} \emph {et~al.} (\bibinfo {collaboration} {MiniBooNE Collaboration}),\ }\href {https://doi.org/10.1103/PhysRevD.103.052002} {\bibfield  {journal} {\bibinfo  {journal} {Phys. Rev. D}\ }\textbf {\bibinfo {volume} {103}},\ \bibinfo {pages} {052002} (\bibinfo {year} {2021})}\BibitemShut {NoStop}%
\bibitem [{\citenamefont {Kelly}\ and\ \citenamefont {Kopp}(2023)}]{Kelly:2022uaa}%
  \BibitemOpen
  \bibfield  {author} {\bibinfo {author} {\bibfnamefont {K.~J.}\ \bibnamefont {Kelly}}\ and\ \bibinfo {author} {\bibfnamefont {J.}~\bibnamefont {Kopp}},\ }\href {https://doi.org/10.1007/JHEP05(2023)113} {\bibfield  {journal} {\bibinfo  {journal} {{JHEP}}\ }\textbf {\bibinfo {volume} {2023}},\ \bibinfo {pages} {113} (\bibinfo {year} {2023})}\BibitemShut {NoStop}%
\bibitem [{\citenamefont {Brdar}\ and\ \citenamefont {Kopp}(2022)}]{Brdar_2022}%
  \BibitemOpen
  \bibfield  {author} {\bibinfo {author} {\bibfnamefont {V.}~\bibnamefont {Brdar}}\ and\ \bibinfo {author} {\bibfnamefont {J.}~\bibnamefont {Kopp}},\ }\href {https://doi.org/10.1103/PhysRevD.105.115024} {\bibfield  {journal} {\bibinfo  {journal} {Phys. Rev. D}\ }\textbf {\bibinfo {volume} {105}},\ \bibinfo {pages} {115024} (\bibinfo {year} {2022})}\BibitemShut {NoStop}%
\bibitem [{\citenamefont {Diaz}\ \emph {et~al.}(2020)\citenamefont {Diaz}, \citenamefont {Arg{\"u}elles}, \citenamefont {Collin}, \citenamefont {Conrad},\ and\ \citenamefont {Shaevitz}}]{Diaz:lightsterile}%
  \BibitemOpen
  \bibfield  {author} {\bibinfo {author} {\bibfnamefont {A.}~\bibnamefont {Diaz}}, \bibinfo {author} {\bibfnamefont {C.}~\bibnamefont {Arg{\"u}elles}}, \bibinfo {author} {\bibfnamefont {G.}~\bibnamefont {Collin}}, \bibinfo {author} {\bibfnamefont {J.}~\bibnamefont {Conrad}},\ and\ \bibinfo {author} {\bibfnamefont {M.}~\bibnamefont {Shaevitz}},\ }\href {https://doi.org/https://doi.org/10.1016/j.physrep.2020.08.005} {\bibfield  {journal} {\bibinfo  {journal} {Phys. Rep.}\ }\textbf {\bibinfo {volume} {884}},\ \bibinfo {pages} {1} (\bibinfo {year} {2020})}\BibitemShut {NoStop}%
\bibitem [{\citenamefont {B{\"o}ser}\ \emph {et~al.}(2020)\citenamefont {B{\"o}ser}, \citenamefont {Buck}, \citenamefont {Giunti}, \citenamefont {Lesgourgues}, \citenamefont {Ludhova}, \citenamefont {Mertens}, \citenamefont {Schukraft},\ and\ \citenamefont {Wurm}}]{Sebas:lightsterile}%
  \BibitemOpen
  \bibfield  {author} {\bibinfo {author} {\bibfnamefont {S.}~\bibnamefont {B{\"o}ser}}, \bibinfo {author} {\bibfnamefont {C.}~\bibnamefont {Buck}}, \bibinfo {author} {\bibfnamefont {C.}~\bibnamefont {Giunti}}, \bibinfo {author} {\bibfnamefont {J.}~\bibnamefont {Lesgourgues}}, \bibinfo {author} {\bibfnamefont {L.}~\bibnamefont {Ludhova}}, \bibinfo {author} {\bibfnamefont {S.}~\bibnamefont {Mertens}}, \bibinfo {author} {\bibfnamefont {A.}~\bibnamefont {Schukraft}},\ and\ \bibinfo {author} {\bibfnamefont {M.}~\bibnamefont {Wurm}},\ }\href {https://doi.org/https://doi.org/10.1016/j.ppnp.2019.103736} {\bibfield  {journal} {\bibinfo  {journal} {Prog. Part. Nucl. Phys.}\ }\textbf {\bibinfo {volume} {111}},\ \bibinfo {pages} {103736} (\bibinfo {year} {2020})}\BibitemShut {NoStop}%
\bibitem [{\citenamefont {Vergani}\ \emph {et~al.}(2021)\citenamefont {Vergani}, \citenamefont {Kamp}, \citenamefont {Diaz}, \citenamefont {Arg\"uelles}, \citenamefont {Conrad}, \citenamefont {Shaevitz},\ and\ \citenamefont {Uchida}}]{Vergani:2021tgc}%
  \BibitemOpen
  \bibfield  {author} {\bibinfo {author} {\bibfnamefont {S.}~\bibnamefont {Vergani}}, \bibinfo {author} {\bibfnamefont {N.~W.}\ \bibnamefont {Kamp}}, \bibinfo {author} {\bibfnamefont {A.}~\bibnamefont {Diaz}}, \bibinfo {author} {\bibfnamefont {C.~A.}\ \bibnamefont {Arg\"uelles}}, \bibinfo {author} {\bibfnamefont {J.~M.}\ \bibnamefont {Conrad}}, \bibinfo {author} {\bibfnamefont {M.~H.}\ \bibnamefont {Shaevitz}},\ and\ \bibinfo {author} {\bibfnamefont {M.~A.}\ \bibnamefont {Uchida}},\ }\href {https://doi.org/10.1103/PhysRevD.104.095005} {\bibfield  {journal} {\bibinfo  {journal} {Phys. Rev. D}\ }\textbf {\bibinfo {volume} {104}},\ \bibinfo {pages} {095005} (\bibinfo {year} {2021})}\BibitemShut {NoStop}%
\bibitem [{\citenamefont {Alvarez-Ruso}\ and\ \citenamefont {Saul-Sala}(2017)}]{Alvarez-Ruso:2017hdm}%
  \BibitemOpen
  \bibfield  {author} {\bibinfo {author} {\bibfnamefont {L.}~\bibnamefont {Alvarez-Ruso}}\ and\ \bibinfo {author} {\bibfnamefont {E.}~\bibnamefont {Saul-Sala}},\ }in\ \href@noop {} {\emph {\bibinfo {booktitle} {{Prospects in Neutrino Physics}}}}\ (\bibinfo {year} {2017})\ \Eprint {https://arxiv.org/abs/1705.00353} {arXiv:1705.00353 [hep-ph]} \BibitemShut {NoStop}%
\bibitem [{\citenamefont {Fischer}\ \emph {et~al.}(2020)\citenamefont {Fischer}, \citenamefont {Hern\'andez-Cabezudo},\ and\ \citenamefont {Schwetz}}]{Fischer:heavysterile}%
  \BibitemOpen
  \bibfield  {author} {\bibinfo {author} {\bibfnamefont {O.}~\bibnamefont {Fischer}}, \bibinfo {author} {\bibfnamefont {A.}~\bibnamefont {Hern\'andez-Cabezudo}},\ and\ \bibinfo {author} {\bibfnamefont {T.}~\bibnamefont {Schwetz}},\ }\href {https://doi.org/10.1103/PhysRevD.101.075045} {\bibfield  {journal} {\bibinfo  {journal} {Phys. Rev. D}\ }\textbf {\bibinfo {volume} {101}},\ \bibinfo {pages} {075045} (\bibinfo {year} {2020})}\BibitemShut {NoStop}%
\bibitem [{\citenamefont {Abdullahi}\ \emph {et~al.}(2021)\citenamefont {Abdullahi}, \citenamefont {Hostert},\ and\ \citenamefont {Pascoli}}]{Abdullahi:2020nyr}%
  \BibitemOpen
  \bibfield  {author} {\bibinfo {author} {\bibfnamefont {A.}~\bibnamefont {Abdullahi}}, \bibinfo {author} {\bibfnamefont {M.}~\bibnamefont {Hostert}},\ and\ \bibinfo {author} {\bibfnamefont {S.}~\bibnamefont {Pascoli}},\ }\href {https://doi.org/10.1016/j.physletb.2021.136531} {\bibfield  {journal} {\bibinfo  {journal} {Phys. Lett. B}\ }\textbf {\bibinfo {volume} {820}},\ \bibinfo {pages} {136531} (\bibinfo {year} {2021})}\BibitemShut {NoStop}%
\bibitem [{\citenamefont {Bertuzzo}\ \emph {et~al.}(2018)\citenamefont {Bertuzzo}, \citenamefont {Jana}, \citenamefont {Machado},\ and\ \citenamefont {Zukanovich~Funchal}}]{Bertuzzo:2018itn}%
  \BibitemOpen
  \bibfield  {author} {\bibinfo {author} {\bibfnamefont {E.}~\bibnamefont {Bertuzzo}}, \bibinfo {author} {\bibfnamefont {S.}~\bibnamefont {Jana}}, \bibinfo {author} {\bibfnamefont {P.~A.~N.}\ \bibnamefont {Machado}},\ and\ \bibinfo {author} {\bibfnamefont {R.}~\bibnamefont {Zukanovich~Funchal}},\ }\href {https://doi.org/10.1103/PhysRevLett.121.241801} {\bibfield  {journal} {\bibinfo  {journal} {Phys. Rev. Lett.}\ }\textbf {\bibinfo {volume} {121}},\ \bibinfo {pages} {241801} (\bibinfo {year} {2018})}\BibitemShut {NoStop}%
\bibitem [{\citenamefont {Ballett}\ \emph {et~al.}(2019)\citenamefont {Ballett}, \citenamefont {Pascoli},\ and\ \citenamefont {Ross-Lonergan}}]{Ballett:2018ynz}%
  \BibitemOpen
  \bibfield  {author} {\bibinfo {author} {\bibfnamefont {P.}~\bibnamefont {Ballett}}, \bibinfo {author} {\bibfnamefont {S.}~\bibnamefont {Pascoli}},\ and\ \bibinfo {author} {\bibfnamefont {M.}~\bibnamefont {Ross-Lonergan}},\ }\href {https://doi.org/10.1103/PhysRevD.99.071701} {\bibfield  {journal} {\bibinfo  {journal} {Phys. Rev. D}\ }\textbf {\bibinfo {volume} {99}},\ \bibinfo {pages} {071701} (\bibinfo {year} {2019})}\BibitemShut {NoStop}%
\bibitem [{\citenamefont {Abratenko}\ \emph {et~al.}(2022{\natexlab{f}})\citenamefont {Abratenko} \emph {et~al.}}]{GenieUBTune}%
  \BibitemOpen
  \bibfield  {author} {\bibinfo {author} {\bibfnamefont {P.}~\bibnamefont {Abratenko}} \emph {et~al.} (\bibinfo {collaboration} {MicroBooNE Collaboration}),\ }\href {https://doi.org/10.1103/PhysRevD.105.072001} {\bibfield  {journal} {\bibinfo  {journal} {Phys. Rev. D}\ }\textbf {\bibinfo {volume} {105}},\ \bibinfo {pages} {072001} (\bibinfo {year} {2022}{\natexlab{f}})}\BibitemShut {NoStop}%
\bibitem [{\citenamefont {Allison}\ \emph {et~al.}(2016)\citenamefont {Allison} \emph {et~al.}}]{ALLISON2016186}%
  \BibitemOpen
  \bibfield  {author} {\bibinfo {author} {\bibfnamefont {J.}~\bibnamefont {Allison}} \emph {et~al.},\ }\href {https://doi.org/https://doi.org/10.1016/j.nima.2016.06.125} {\bibfield  {journal} {\bibinfo  {journal} {{Nucl. Instrum. Meth. A}}\ }\textbf {\bibinfo {volume} {835}},\ \bibinfo {pages} {186} (\bibinfo {year} {2016})}\BibitemShut {NoStop}%
\bibitem [{\citenamefont {Allison}\ \emph {et~al.}(2006)\citenamefont {Allison} \emph {et~al.}}]{Geant:1610988}%
  \BibitemOpen
  \bibfield  {author} {\bibinfo {author} {\bibfnamefont {J.}~\bibnamefont {Allison}} \emph {et~al.},\ }\href {https://doi.org/10.1109/TNS.2006.869826} {\bibfield  {journal} {\bibinfo  {journal} {{IEEE Trans. Nucl. Sci.}}\ }\textbf {\bibinfo {volume} {53}},\ \bibinfo {pages} {270} (\bibinfo {year} {2006})}\BibitemShut {NoStop}%
\bibitem [{\citenamefont {Agostinelli}\ \emph {et~al.}(2003)\citenamefont {Agostinelli} \emph {et~al.}}]{AGOSTINELLI2003250}%
  \BibitemOpen
  \bibfield  {author} {\bibinfo {author} {\bibfnamefont {S.}~\bibnamefont {Agostinelli}} \emph {et~al.},\ }\href {https://doi.org/https://doi.org/10.1016/S0168-9002(03)01368-8} {\bibfield  {journal} {\bibinfo  {journal} {{Nucl. Instrum. Meth. A}}\ }\textbf {\bibinfo {volume} {506}},\ \bibinfo {pages} {250} (\bibinfo {year} {2003})}\BibitemShut {NoStop}%
\bibitem [{\citenamefont {Acciarri}\ \emph {et~al.}(2018)\citenamefont {Acciarri} \emph {et~al.}}]{pandora}%
  \BibitemOpen
  \bibfield  {author} {\bibinfo {author} {\bibfnamefont {R.}~\bibnamefont {Acciarri}} \emph {et~al.} (\bibinfo {collaboration} {MicroBooNE Collaboration}),\ }\href {https://doi.org/10.1140/epjc/s10052-017-5481-6} {\bibfield  {journal} {\bibinfo  {journal} {Eur. Phys. J. C}\ }\textbf {\bibinfo {volume} {78}},\ \bibinfo {pages} {82} (\bibinfo {year} {2018})}\BibitemShut {NoStop}%
\bibitem [{\citenamefont {Adams}\ \emph {et~al.}(2019)\citenamefont {Adams} \emph {et~al.}}]{bib:CRT}%
  \BibitemOpen
  \bibfield  {author} {\bibinfo {author} {\bibfnamefont {C.}~\bibnamefont {Adams}} \emph {et~al.} (\bibinfo {collaboration} {MicroBooNE Collaboration}),\ }\href {https://doi.org/10.1088/1748-0221/14/04/P04004} {\bibfield  {journal} {\bibinfo  {journal} {{J. Instrum.}}\ }\textbf {\bibinfo {volume} {14}},\ \bibinfo {pages} {P04004} (\bibinfo {year} {2019})}\BibitemShut {NoStop}%
\bibitem [{\citenamefont {Adams}\ \emph {et~al.}(2020)\citenamefont {Adams} \emph {et~al.}}]{Adams_2020}%
  \BibitemOpen
  \bibfield  {author} {\bibinfo {author} {\bibfnamefont {C.}~\bibnamefont {Adams}} \emph {et~al.} (\bibinfo {collaboration} {MicroBooNE Collaboration}),\ }\href {https://doi.org/10.1088/1748-0221/15/02/P02007} {\bibfield  {journal} {\bibinfo  {journal} {{J. Instrum.}}\ }\textbf {\bibinfo {volume} {15}},\ \bibinfo {pages} {P02007} (\bibinfo {year} {2020})}\BibitemShut {NoStop}%
\bibitem [{\citenamefont {Berger}\ \emph {et~al.}(2017)\citenamefont {Berger}, \citenamefont {Coursey}, \citenamefont {Zucker},\ and\ \citenamefont {Chang}}]{nist_stopping_power}%
  \BibitemOpen
  \bibfield  {author} {\bibinfo {author} {\bibfnamefont {M.~J.}\ \bibnamefont {Berger}}, \bibinfo {author} {\bibfnamefont {J.~S.}\ \bibnamefont {Coursey}}, \bibinfo {author} {\bibfnamefont {M.~A.}\ \bibnamefont {Zucker}},\ and\ \bibinfo {author} {\bibfnamefont {J.}~\bibnamefont {Chang}},\ }\href {https://dx.doi.org/10.18434/T4NC7P} {\emph {\bibinfo {title} {{Stopping-Power \& Range Tables for Electrons, Protons, and Helium Ions}}}},\ \bibinfo {series} {NIST Standard Reference Database}\ No.\ \bibinfo {number} {124}\ (\bibinfo  {publisher} {National Institute of Standards and Technology},\ \bibinfo {year} {2017})\BibitemShut {NoStop}%
\bibitem [{\citenamefont {Groom}\ \emph {et~al.}(2001)\citenamefont {Groom}, \citenamefont {Mokhov},\ and\ \citenamefont {Striganov}}]{pdg_muon_stopping_power}%
  \BibitemOpen
  \bibfield  {author} {\bibinfo {author} {\bibfnamefont {D.~E.}\ \bibnamefont {Groom}}, \bibinfo {author} {\bibfnamefont {N.~V.}\ \bibnamefont {Mokhov}},\ and\ \bibinfo {author} {\bibfnamefont {S.~I.}\ \bibnamefont {Striganov}},\ }\href {https://doi.org/https://doi.org/10.1006/adnd.2001.0861} {\bibfield  {journal} {\bibinfo  {journal} {Atomic Data and Nuclear Data Tables}\ }\textbf {\bibinfo {volume} {78}},\ \bibinfo {pages} {183} (\bibinfo {year} {2001})}\BibitemShut {NoStop}%
\bibitem [{\citenamefont {Abratenko}\ \emph {et~al.}(2021)\citenamefont {Abratenko} \emph {et~al.}}]{MicroBooNE:2021ddy}%
  \BibitemOpen
  \bibfield  {author} {\bibinfo {author} {\bibfnamefont {P.}~\bibnamefont {Abratenko}} \emph {et~al.} (\bibinfo {collaboration} {MicroBooNE}),\ }\href {https://doi.org/10.1007/JHEP12(2021)153} {\bibfield  {journal} {\bibinfo  {journal} {{JHEP}}\ }\textbf {\bibinfo {volume} {12}},\ \bibinfo {pages} {153} (\bibinfo {year} {2021})}\BibitemShut {NoStop}%
\bibitem [{\citenamefont {Abratenko}\ \emph {et~al.}(2022{\natexlab{g}})\citenamefont {Abratenko} \emph {et~al.}}]{MicroBooNE:2021roa}%
  \BibitemOpen
  \bibfield  {author} {\bibinfo {author} {\bibfnamefont {P.}~\bibnamefont {Abratenko}} \emph {et~al.} (\bibinfo {collaboration} {MicroBooNE Collaboration}),\ }\href {https://doi.org/10.1140/epjc/s10052-022-10270-8} {\bibfield  {journal} {\bibinfo  {journal} {Eur. Phys. J. C}\ }\textbf {\bibinfo {volume} {82}},\ \bibinfo {pages} {454} (\bibinfo {year} {2022}{\natexlab{g}})}\BibitemShut {NoStop}%
\bibitem [{Sup()}]{SupplementalMaterial}%
  \BibitemOpen
  \href@noop {} {}\bibinfo {note} {See Supplemental Material, which includes Refs. \cite{bib:CRT,MicroBooNE:2021wad,bib:LEEmodel}, for additional information for this analysis, including the impact of the CRT on selection, constraint covariance and its impact, smoothing of detector systematics, data validation, updates to the signal model, a summary of predicted and observed event rates, the complete results of the two-hypothesis tests, a summary of uncertainties, and event displays for selected candidate events.}\BibitemShut {Stop}%
\bibitem [{\citenamefont {Eaton}(1983)}]{Eaton1983-xi}%
  \BibitemOpen
  \bibfield  {author} {\bibinfo {author} {\bibfnamefont {M.~L.}\ \bibnamefont {Eaton}},\ }\href@noop {} {\emph {\bibinfo {title} {Multivariate statistics}}},\ Probability \& Mathematical Statistics\ (\bibinfo  {publisher} {John Wiley \& Sons},\ \bibinfo {address} {Nashville, TN},\ \bibinfo {year} {1983})\ pp.\ \bibinfo {pages} {116--117}\BibitemShut {NoStop}%
\bibitem [{\citenamefont {Ji}\ \emph {et~al.}(2020)\citenamefont {Ji}, \citenamefont {Gu}, \citenamefont {Qian}, \citenamefont {Wei},\ and\ \citenamefont {Zhang}}]{Ji:2019yca}%
  \BibitemOpen
  \bibfield  {author} {\bibinfo {author} {\bibfnamefont {X.}~\bibnamefont {Ji}}, \bibinfo {author} {\bibfnamefont {W.}~\bibnamefont {Gu}}, \bibinfo {author} {\bibfnamefont {X.}~\bibnamefont {Qian}}, \bibinfo {author} {\bibfnamefont {H.}~\bibnamefont {Wei}},\ and\ \bibinfo {author} {\bibfnamefont {C.}~\bibnamefont {Zhang}},\ }\href {https://doi.org/10.1016/j.nima.2020.163677} {\bibfield  {journal} {\bibinfo  {journal} {Nucl. Instrum. Meth. A}\ }\textbf {\bibinfo {volume} {961}},\ \bibinfo {pages} {163677} (\bibinfo {year} {2020})}\BibitemShut {NoStop}%
\bibitem [{\citenamefont {Aguilar-Arevalo}\ \emph {et~al.}(2007)\citenamefont {Aguilar-Arevalo} \emph {et~al.}}]{PhysRevLett.98.231801}%
  \BibitemOpen
  \bibfield  {author} {\bibinfo {author} {\bibfnamefont {A.~A.}\ \bibnamefont {Aguilar-Arevalo}} \emph {et~al.} (\bibinfo {collaboration} {MiniBooNE Collaboration}),\ }\href {https://doi.org/10.1103/PhysRevLett.98.231801} {\bibfield  {journal} {\bibinfo  {journal} {Phys. Rev. Lett.}\ }\textbf {\bibinfo {volume} {98}},\ \bibinfo {pages} {231801} (\bibinfo {year} {2007})}\BibitemShut {NoStop}%
\bibitem [{\citenamefont {Abratenko}\ \emph {et~al.}(2022{\natexlab{h}})\citenamefont {Abratenko} \emph {et~al.}}]{MicroBooNE:2022tdd}%
  \BibitemOpen
  \bibfield  {author} {\bibinfo {author} {\bibfnamefont {P.}~\bibnamefont {Abratenko}} \emph {et~al.} (\bibinfo {collaboration} {MicroBooNE Collaboration}),\ }\href {https://doi.org/10.1103/PhysRevD.106.L051102} {\bibfield  {journal} {\bibinfo  {journal} {Phys. Rev. D}\ }\textbf {\bibinfo {volume} {106}},\ \bibinfo {pages} {L051102} (\bibinfo {year} {2022}{\natexlab{h}})}\BibitemShut {NoStop}%
\bibitem [{\citenamefont {Feldman}\ and\ \citenamefont {Cousins}(1998)}]{Feldman:1997qc}%
  \BibitemOpen
  \bibfield  {author} {\bibinfo {author} {\bibfnamefont {G.~J.}\ \bibnamefont {Feldman}}\ and\ \bibinfo {author} {\bibfnamefont {R.~D.}\ \bibnamefont {Cousins}},\ }\href {https://doi.org/10.1103/PhysRevD.57.3873} {\bibfield  {journal} {\bibinfo  {journal} {Phys. Rev. D}\ }\textbf {\bibinfo {volume} {57}},\ \bibinfo {pages} {3873} (\bibinfo {year} {1998})}\BibitemShut {NoStop}%
\bibitem [{\citenamefont {Abratenko}\ \emph {et~al.}(2022{\natexlab{i}})\citenamefont {Abratenko} \emph {et~al.}}]{PhysRevD.105.L051102}%
  \BibitemOpen
  \bibfield  {author} {\bibinfo {author} {\bibfnamefont {P.}~\bibnamefont {Abratenko}} \emph {et~al.} (\bibinfo {collaboration} {MicroBooNE Collaboration}),\ }\href {https://doi.org/10.1103/PhysRevD.105.L051102} {\bibfield  {journal} {\bibinfo  {journal} {Phys. Rev. D}\ }\textbf {\bibinfo {volume} {105}},\ \bibinfo {pages} {L051102} (\bibinfo {year} {2022}{\natexlab{i}})}\BibitemShut {NoStop}%
\bibitem [{\citenamefont {Abratenko}\ \emph {et~al.}(2023)\citenamefont {Abratenko} \emph {et~al.}}]{MicroBooNE:2022sdp}%
  \BibitemOpen
  \bibfield  {author} {\bibinfo {author} {\bibfnamefont {P.}~\bibnamefont {Abratenko}} \emph {et~al.} (\bibinfo {collaboration} {MicroBooNE Collaboration}),\ }\href {https://doi.org/10.1103/PhysRevLett.130.011801} {\bibfield  {journal} {\bibinfo  {journal} {Phys. Rev. Lett.}\ }\textbf {\bibinfo {volume} {130}},\ \bibinfo {pages} {011801} (\bibinfo {year} {2023})}\BibitemShut {NoStop}%
\bibitem [{\citenamefont {Junk}(1999)}]{CLs_method}%
  \BibitemOpen
  \bibfield  {author} {\bibinfo {author} {\bibfnamefont {T.}~\bibnamefont {Junk}},\ }\href {https://doi.org/https://doi.org/10.1016/S0168-9002(99)00498-2} {\bibfield  {journal} {\bibinfo  {journal} {{Nucl. Instrum. Meth. A}}\ }\textbf {\bibinfo {volume} {434}},\ \bibinfo {pages} {435} (\bibinfo {year} {1999})}\BibitemShut {NoStop}%
\bibitem [{\citenamefont {Read}(2000)}]{Read:451614}%
  \BibitemOpen
  \bibfield  {author} {\bibinfo {author} {\bibfnamefont {A.~L.}\ \bibnamefont {Read}}\ }\href {https://doi.org/10.5170/CERN-2000-005.81} {10.5170/CERN-2000-005.81} (\bibinfo {year} {2000})\BibitemShut {NoStop}%
\bibitem [{bib()}]{bib:LEEmodel}%
  \BibitemOpen
  \href@noop {} {}\bibinfo {note} {MicroBooNE Collaboration, \href{https://microboone.fnal.gov/wp-content/uploads/MICROBOONE-NOTE-1043-PUB.pdf}{MICROBOONE-NOTE-1043-PUB}, DOI:\href{https://doi.org/10.2172/1573217}{10.2172/1573217} (2018)}\BibitemShut {NoStop}%
\end{thebibliography}%

\end{document}